\documentstyle [epsfig,12pt]{article}
\clearpage
\newcommand{\bz}{\mbox{$\mathrm{B}^{0}$}}
\newcommand{\bd}{\mbox{$\mathrm{B}^{0}_{d}$}}
\newcommand{\bs}{\mbox{$\mathrm{B}^{0}_{s}$}}
\newcommand{\dms}{\mbox{$\Delta{m_{s}}$}}
\newcommand{\dmd}{\mbox{$\Delta{m_{d}}$}}
\newcommand{\ps}{\mbox{$\mathrm{ps}^{-1}$}}

\begin{document}
\newcommand{\ev}[1]{\langle #1 \rangle}
\thispagestyle{empty}

\vspace{1.0 cm}
\begin{flushright}
{\bf LAL 98-76}\\
October 1998 \\
\vspace{1.0 cm}

\end{flushright}
\vspace{2.5 cm}
\begin{center}

{\Large\bf Oscillations of neutral B mesons systems
}
\vspace{1.5cm}

{\large J. Boucrot\\
Laboratoire de l'Acc\'el\'erateur Lin\'eaire,\\
IN2P3-CNRS et Universit\'e de Paris-Sud,\\
B.P. 34,  91898 Orsay CEDEX, France
}

\end{center}
\vspace{3.0 cm}

\begin{abstract}

 The oscillation phenomenon in the neutral B mesons 
 systems is now well established.
 The motivations and principles of the measurements are given; then 
 the most recent results from the LEP experiments, the CDF
 collaboration
 at Fermilab and the SLD collaboration at SLAC are reviewed.
 The present world average of the
 $\bd$ meson oscillation frequency is $\dmd = 0.471 \pm 0.016 \, \ps$
 and the lower limit on the $\bs$ oscillation frequency is
 $\dms > 12.4 \, \ps \mbox{ at 95\% CL} $.

\end{abstract}

\vspace*{2.0 cm}

\begin{center}
\it{Review talk given at the\\
Third International Conference on Hyperons, Charm and Beauty Hadrons,\\
Genoa, Italy, 29 June - 3 July 1998}
\end{center}

\newpage
 
\section{Physics motivations}

It is known since a long time that $\bd ( d \overline{b})$  and $\bs ( s
\overline{b})$ mesons can undergo particle-antiparticle
mixing, due to second-order weak interactions which involve box diagrams. If CP
violation is neglected, the CP eigenstates of the
$\bz  - \overline{\bz}$ system are also mass eigenstates and the probability
that a purely $\bz$ state produced at time t=0
will decay as a $\bz$ or $\overline{\bz}$  at time t is:

\begin{equation}
p(t,\mu) = \frac{ e^{-(t/\tau)}\, [1+\mu cos(\Delta{m} t)]}{2\tau}
\end{equation}

The oscillation frequency $\Delta{m}$ is the mass
difference between the two mass eigenstates,
 $\tau$ is the average lifetime
and $\mu$ = +1 (resp -1) if the final state is a $\bz$ (resp $\overline{\bz}$).

In the framework of the Standard Model, $\Delta{m}$ may be computed 
with a rather large uncertainty coming from QCD correction factors.
 The ratio of the oscillation frequencies is under better control:

\begin{equation}
\frac{\dms}{\dmd} = \xi^{2}   
                      \frac{m_{B_{s}}}{m_{B_{d}}}
                      \left| \frac{V_{ts}}{V_{td}} \right|^{2}  
\end{equation}

where $\xi = 1.16 \pm 0.10$ contains all  theoretical uncertainties.
 Therefore the measurement of 
$\bd$ and $\bs$ oscillation frequencies should give constraints on
the poorly known CKM matrix elements
$V_{ts}$ and $V_{td}$. These constraints are conveniently expressed in the framework of the
Wolfenstein parametrisation of the CKM matrix, introducing the so-called Unitarity Triangle.

\section{Experiments and data}

\subsection{LEP experiments}

The four LEP experiments ALEPH, DELPHI, L3 and OPAL use data taken at the Z
peak from 1991 to 1995.
They have accumulated about 4 million hadronic Z decays each, i.e. about one
million pairs of B mesons,
with the following properties:

- B mesons are produced with a large boost (mean momentum around 35 GeV/c),
  giving a decay length around 2 to 3 mm;

 - All experiments have good Silicon vertex detectors and obtain a resolution on
   the decay length around 300 micrometers;

 - The particle identification, which is very important for $\bs$ studies, is
   done through dE/dx measurements in ALEPH
 and OPAL, and using  a RICH (Ring Imaging Cerenkov counter) in DELPHI.

\subsection{CDF}

    The Collider Detector Facility at Fermilab has collected 130
$\mathrm{pb}^{-1}$ of $p \overline{p}$ collisions at
centre of mass energy 1.8 TeV since 1992. At the Collider there is a huge B
meson production rate, with a huge background.
The B mesons are produced with a boost which is less favourable than at LEP/SLC
(mean momentum around 15 GeV/c).

\subsection{SLD}

 The SLD collaboration at SLAC has taken around 500000 hadronic Z0 decays
 between
1993 and 1997, 350000 of which are analyzed.
 The excellent performances
of the SLD vertex detector allow a much better decay length
 resolution than LEP experiments (around 50-100 micrometers). In addition, the
 electron beam polarisation may be used as a powerful
 tool for the B meson tagging \cite{sldbs}.
 The particle identification is done using a Ring
 Imaging Cerenkov counter.

\section{Time-dependent oscillation measurements}

All analyses require reconstruction of the decay time of the $\bz  $ mesons. This 
is done by the measurement of the  decay length and of the  momentum
of the $\bz  $.
As it is impossible to describe in detail all existing measurements, only the
basic principles are given in this section. 

\subsection{$\bz$ meson identification}  

The $\bz  $ meson is in general identified by the detection of a high energy,
high Pt lepton. 

Inclusive lepton analyses demand that this lepton has a large 
impact parameter with respect to the interaction point. This                   
gives several tens of thousands of events, but with 
a low purity (e.g. 10\% for the $\bs$).

 In addition, one can search  for
the exclusive decay of a charmed meson. For instance for $\bd$ decays one
looks for a $D^{*-} l^{+}$ final state, with $D^{*-} \rightarrow D^{0} \pi^{-}$;
the $D^{0}$ meson is then searched for via its hadronic decays e.g. $K^{-} 
\pi^{+}$.
For $\bs$ decays one looks for a $D_{s}^{-} l^{+}$ final state, then for 
$D_{s}$ hadronic decays such as $\phi \pi$.
The charmed meson is used together with the lepton to reconstruct the $\bz  $ 
decay vertex; then the decay length is computed from this vertex and the event 
interaction point. These semi-exclusive analyses give some hundreds of events
per experiment, with a purity which may be as high as 50\%.

In a recent analysis DELPHI \cite{delb0s}
reconstructs fully hadronic decays of the $\bs$.
This gives some tens of events, with a very high purity.

\subsection{Decay time measurement and error}

The $\bz  $ boost is computed using the lepton, the reconstructed charmed meson
and, when possible, the neutrino energy estimated from the missing energy.
Typical values of the decay time resolution for LEP experiments are around 0.3 ps for 
inclusive lepton analyses, 0.15 ps for charm-lepton analyses, and 0.06 ps for 
the DELPHI exclusive $\bs$ reconstruction. SLD obtains a resolution of 0.06 ps
for $D_{s} l$ events. 

\subsection{Tagging of the $\bz  $ at decay}

One has to know if the detected $\bz  $ meson has oscillated or not before
decaying; therefore it is necessary to tag it both at production and at
decay time.

The tagging of the $\bz  $ at decay time is obtained simply  by the charge
 of the  
energetic lepton, since e.g. a positive lepton $l^{+}$ is the signature of the
decay of a $\bz  $ meson. This final state tagging is not perfect since one 
may detect a lepton from a $ b \rightarrow c \rightarrow l$ decay;
the corresponding mistag probability is computed from simulated events.
 
\subsection{Tagging of the $\bz  $ at production}

This is much more complicated. Usually one divides the event in two opposite
hemispheres defined with respect to the event thrust axis; several tags
 are used in each hemisphere:

\begin{itemize}  
\item In the $\bz  $ hemisphere,
the nature and the charge of the most energetic fragmentation track 
carries information on the production state of the $\bz  $ meson: e.g. an
energetic $K^+$ is produced in association with a $\bs$. 
One can also use the sign of the jet charge in this hemisphere.

\item In the hemisphere opposite to the detected $\bz  $, several tags may be
used. First, one can search for an energetic lepton with large Pt
and impact parameter. One can also use the jet charge of the
hemisphere. The sign of the lepton and of the jet charge indicate if a
$b$ or $\overline b$ quark was produced in this hemisphere.
 
\end{itemize}

This initial state tagging is far from being perfect. There are
experimental difficulties: particle misidentification, incorrect attribution of
a charged track to the primary vertex, etc.. and possible mistags due to 
physics: the b quark in the opposite hemisphere may give a $\bz  $ meson which
may oscillate before decay, or the detected lepton may come from the decay of a 
charmed hadron. All these mistag probabilities must be carefully computed      
using the simulation.

\section{Results}

There are presently about 35 different time-dependent $\bz  $ oscillation
analyses, some of them being presented for the first time at this
conference. 
 The oscillation frequency is fitted from a
 likelihood fit to the complicated decay probability function
 which includes all the tags, mistag probabilities 
and efficiencies coming from detailed Monte Carlo studies.

\subsection{Measurement of \dmd}

Four new measurements are available since the 1997 Summer conferences:

- one from CDF \cite{dmdcdf}, using $D l$ and $D^{*} l$  correlations
with the following result:

\begin{equation}
\dmd = 0.471 ^{+0.078}_{-0.068} \mathrm{(stat)} \pm 0.034 \mathrm{(syst)}
  \quad  \ps   
\end{equation}

- from L3 \cite{dmdl3}, three analyses using lepton-lepton decay
  length, lepton-jet charge and lepton-lepton impact parameter, giving:

\begin{equation}
\dmd = 0.444 \pm 0.028 \mathrm{(stat)} \pm 0.028 \mathrm{(syst)} \quad \ps 
\end{equation}

These new results are combined with all the previously published ones, resulting
in an average of 24 $\dmd$ measurements:

\begin{equation}
\dmd = 0.477 \pm 0.017 \quad \ps  \mbox{ (preliminary)}
\end{equation}

This average may be combined with the time-integrated mixing measurements made
by CLEO and ARGUS at the $\Upsilon_{4s}$ resonance, giving the following
world average:
\begin{equation}
\dmd = 0.471 \pm 0.016  \quad \ps  \mbox{(preliminary)}
\end{equation}

This world average is now dominated by systematics (0.013 \ps systematic error, whereas
 the statistical error is  0.010 \ps,
at the 2\% level). 

Figure 1 gives the average of the measurements of \dmd \, obtained by each
ex\-pe\-ri\-ment.

\subsection{Limits on \dms}

 From Eq. (2), the $\bs$ oscillation frequency is expected to be about 20 times greater than
that of the $\bd$. As the $\bs$ meson production rate is about 1/4 that of the
$\bd$, the search for $\bs$ oscillations is therefore an experimental challenge.
 Up to now, no measurement of \dms \, is available; only lower limits are obtained
from 11 existing analyses.

At this conference, the results described below have been presented for the first time.
All limits given below are 95\% Confidence Level limits.

- DELPHI \cite{dmsdel}
has presented a new limit using 280 $ D_{s} l$ events, the $D_{s}$ being
reconstructed in 6 hadronic and 2 semileptonic decay modes. 
They obtain a lower limit of \dms $ > 7.4 \ps$.

- DELPHI \cite{delb0s} gives a lower limit of  \dms $ > 2.4 \ps$
 from  fully reconstructed $\bs$ mesons (they have
$17 \pm 8$ events reconstructed in 8 decay channels). These events give an important contribution
for high \dms  values (above 10 \ps) due to their excellent proper 
time resolution. 

- SLD   \cite{sldbs} has presented 
three analyses: lepton - D , charged dipole, 
lepton + tracks. They exclude the
following regions: 

\begin{equation}
\dms < 1.3 \, \ps \; and \;    2.7 < \dms < 5.3 \, \ps
\end{equation}
                            
- CDF \cite{cdfbs} has presented 
 $1068 \pm 70$ events with  $\phi l$ correlations, from which
they  obtain a lower limit of $\dms > 5.8 \ps$.

At the time of the conference, the overall combination could  not yet be done.
 This has been done since that time by the LEP Working Group,
using the method defined in \cite{refamp}.
The resulting 
 preliminary world average of the amplitudes is displayed in Figure  2,
 and the  pre\-li\-mi\-nary lower limit is \cite{lepwg}:

\begin{equation}
\dms > 12.4 \, \ps \, \mbox{at 95\% CL}
\end{equation}
 with a sensitivity of $13.8 \, \ps$.

\section{Consequences on the Unitarity Triangle}

F. Parodi et al. \cite{ckm} have studied the consequences of the above results for the Unitarity Triangle.
They find rather strong constraints on the triangle apex position:\linebreak
$ \rho = 0.189 \pm 0.074$ , $ \eta = 0.354 \pm 0.045$ 
and also a strong constraint on the $\beta$ angle: $sin(2\beta) = 0.73 \pm 0.08$. 
This is especially interesting before the starting up of the B-factories at SLAC and KEK,
which will measure this angle directly. 

They find also that
within the present know\-ledge of all parameters, the allowed range of $\dms$ is $ 6 < \dms < 21 \ps$    
at 95\% CL. This means that half of the allowed range is already experimentally excluded.

\section{Future prospects}

The most recent individual measurements of the $\bd$ oscillation frequency
from L3 and CDF
are impressively precise. However the present world average has a precision
around 3\%, and therefore future improvements are expected to be modest.

The limit on $\dms$ should improve appreciably in the near future. The LEP
experiments still have room for improvement, especially ALEPH and OPAL.
 SLD should also give new limits using their
full statistics on tape and CDF will give more results on their present data
 using new channels.
 
In a more distant future (2 years), new results should
come from SLD if they run in 1999 (expected sensitivity around
 20 $\ps$). In year 2000, the HERA B experiment at DESY will have data, and at
 the Fermilab Collider the CDF and D0 collaborations will collect new 
data with
 an expected sensitivity of 25 $\ps$. This would explore the full domain allowed
 to $\dms$ by the Standard Model.

\newpage
\section{Conclusions}

Using the most recent CDF and L3 measurements, the 
$\bd$ oscillation frequency is now measured
with an accuracy of almost 3\%:

\begin{equation}
\dmd = 0.471 \pm 0.016 \, \ps  \mbox{(preliminary)}
\end{equation}

 New analyses 
 on the $\bs$ oscillation
frequency give
the following combined new lower limit:

\begin{equation}
\dms > 12.4 \, \ps \mbox{ at 95\% CL (preliminary)}
\end{equation}

This result excludes already half of the allowed region for $\dms$ in the
Standard Model, and gives efficient constraints on the Unitarity Triangle. 
Improvements on the $\dms$ limit are expected 
 from updated LEP analyses and from
new data at SLD and at the Fermilab Collider.

\vskip 1.0cm

\section{Acknowledgements}

I would like to thank the Third International Conference on Hyperons,
Charm and Beauty Hadrons to have invited me for this talk.

 I also wish to thank 
the various people  who provided me with the results of all
collaborations: O. Schneider (ALEPH), B. Wicklund (CDF), A. Stocchi (DELPHI), V.
Andreev (L3), M. Jimack (OPAL) and S. Willocq (SLD). I also thank warmly 
the LEP Working Group on B oscillations for providing me with Figures 1 and 2, 
 and A. Stocchi for the constraints on the Unitarity Triangle \cite{ckm}.


\newpage
\vskip 1.0cm
\begin{figure}[t]
\begin{center}
\mbox{\epsfig{file=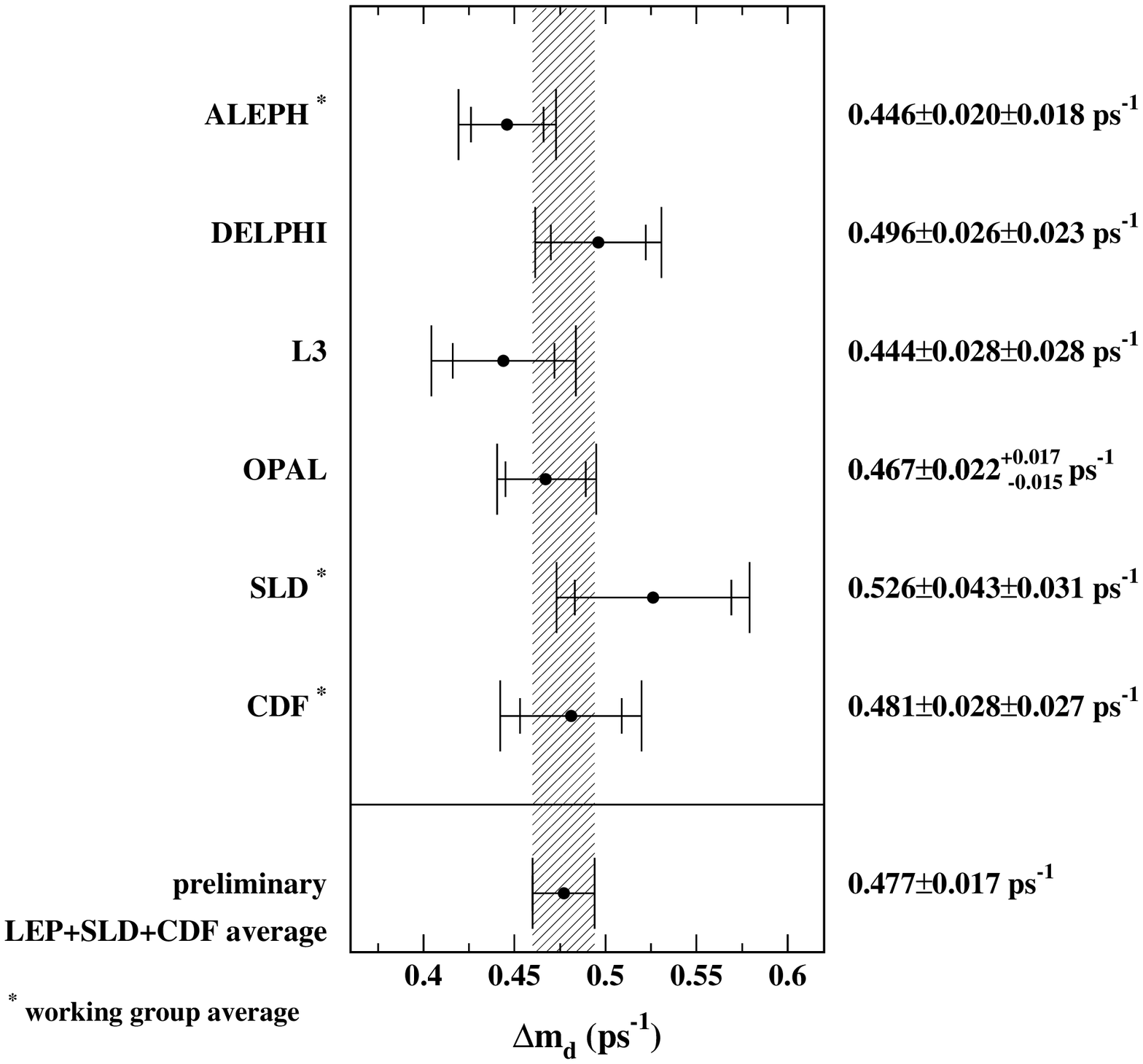, width=15cm,height=17cm}}
\end{center}
\caption{Average $\Delta{m_{d}}$ value per experiment. This figure has been
provided by the LEP Working Group on
B Oscillations [8].}
\label{fig:dmdexp}
\end{figure}

\newpage
\vskip 1.0cm
\begin{figure}[t]
\begin{center}
\mbox{\epsfig{file=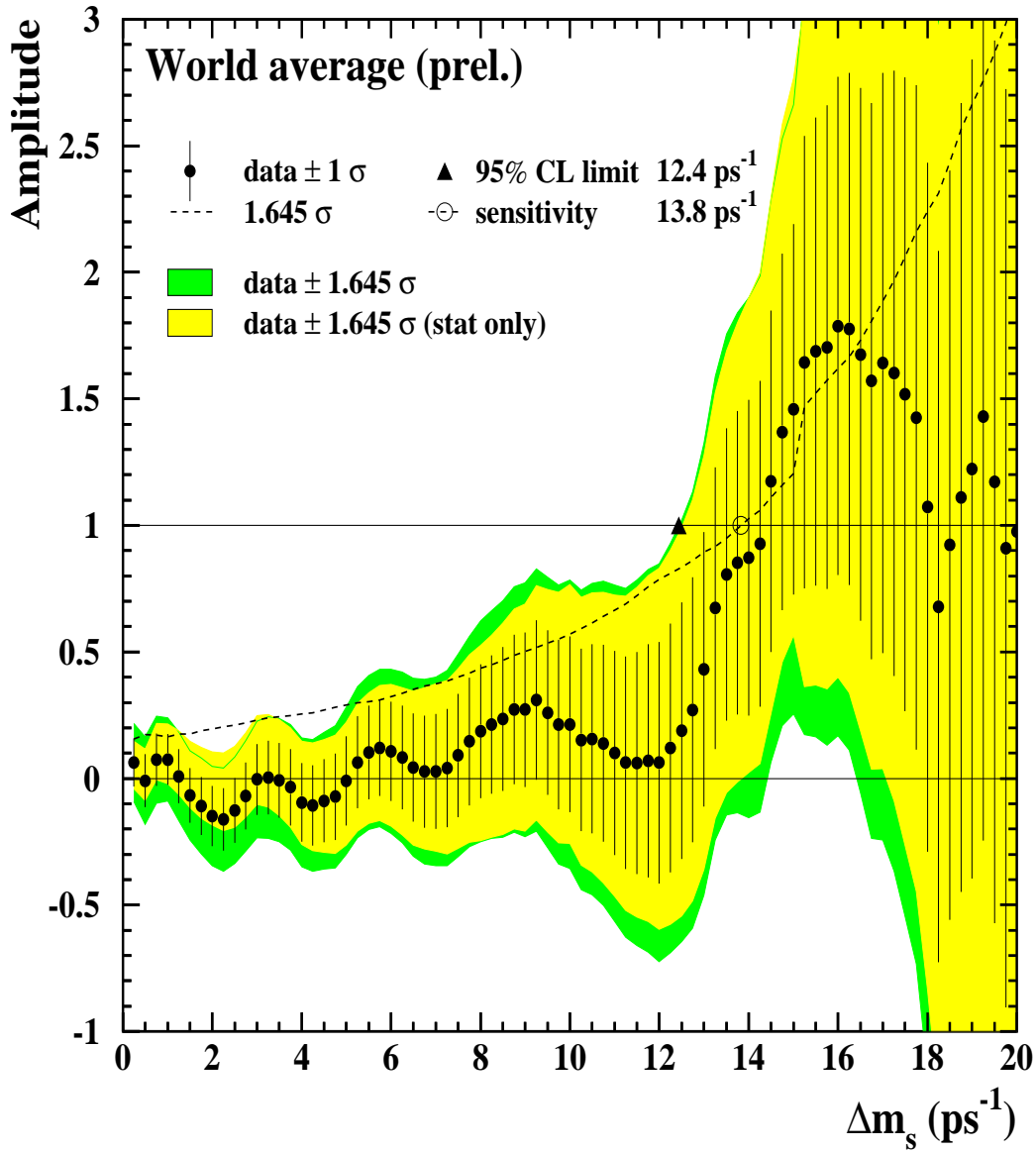,width=15cm, height=17cm}}
\end{center}
\caption{Combined $\mathrm{B}^{0}_{s}$ oscillation amplitude spectrum as a 
function of $\Delta{m_{s}}$. A 95\% CL
lower limit of 12.4 $\mathrm{ps}^{-1}$ on $\Delta{m_{s}}$ is derived from
this spectrum.
All published or preliminary results known as of July 13th 1998 are included in
this figure, provided by the LEP Working Group on B Oscillations [8].}
\label{f:ampbos}
\end{figure}

\end{document}